\documentclass{IEEEcsmagpreprint}

\usepackage[colorlinks,urlcolor=blue,linkcolor=blue,citecolor=blue]{hyperref}
\expandafter\def\expandafter\UrlBreaks\expandafter{\UrlBreaks\do\/\do\*\do\-\do\~\do\'\do\"\do\-}
\usepackage{upmath,color}
\usepackage{comment}
\usepackage{pifont} 

\jvol{XX}
\jnum{XX}
\paper{1}
\jmonth{Feb}
\jname{preprint. SAND2026-16932O} 
\jtitle{preprint. SAND2026-16932O} 
\pubyear{2026}

\begin{document}


\title{The Case of the Mysterious Citations}

\author{Amanda Bienz}
\affil{University of New Mexico, Albuquerque, NM, USA}

\author{Carl Pearson}
\affil{Sandia National Laboratories, Albuquerque, NM, USA}

\author{Simon Garcia de Gonzalo}
\affil{Sandia National Laboratories, Albuquerque, NM, USA}

\markboth{THEME}{THEME}

\begin{abstract}\looseness-1 Mysterious citations are routinely appearing in peer-reviewed publications throughout the scientific community.  In this paper, we developed a semi-automated pipeline and examine the proceedings of four major high-performance computing conferences for both 2021 and 2025, pre- and post-generative AI respectively.  We analyzed each paper for mysterious citations, or those in which the cited title was verified to not exist at the cited location.  Further, we analyzed author correctness throughout citations, flagging citations for which more than half of the cited authors did not match those listed on the cited publication.  While there were zero cases of mysterious citations or incorrect author attributions in any 2021 publication, there were multiple of each in every 2025 conference.  In total, there were 38 mysterious citations and 46 citations for which the majority of listed authors incorrect, spanning all four 2025 proceedings.  No paper within our dataset acknowledged using AI to generate citations even though all four conference policies required acknowledging AI usage, indicating current policies are insufficient.

\end{abstract}

\maketitle

\chapteri{L}arge language models (LLMs) have transitioned in just a few years from research prototypes to widely available tools.
Systems such as ChatGPT~\cite{openai_chatgpt_2022}, Claude~\cite{anthropic_claude_2023}, and Gemini~\cite{google_gemini_2023} are now embedded in everyday workflows, offering fluent text generation, summarization, and synthesis at low cost.
Accessibility, ease of use, and apparent competence have made them attractive assistants for writing tasks in various domains, including academic research~\cite{liang2025quantifying}.

This shift has begun to surface explicitly in the academic publication ecosystem.
Major venues and professional societies, including ACM and IEEE, now publish guidelines on the use of generative AI in paper preparation\cite{acm_authorship_ai_2023, ieee_ai_guidelines_2023}, ranging from disclosure to authorship attribution.
These policies acknowledge a reality: LLMs are already being used in conference and journal submissions, particularly for drafting, editing, and literature review assistance, and their presence is no longer exceptional.

At the same time, the use of LLMs in publications is often not transparent.
Many venues do not require explicit disclosure, and even when policies exist, compliance is uneven.
Anecdotal reports from program committees and reviewers increasingly describe a specific failure mode: ``hallucinated'' or outright fabricated citations inserted by LLMs for related work sections or to substantiate claims made in the publication.  
These references often appear superficially plausible, complete with author lists (featuring names of actual researchers), real venues, and publication dates, but correspond to no real publication. 

This phenomenon poses a burden on the peer-review process. Reviewers cannot reliably detect such errors by inspection alone. Verifying a suspicious reference may require checking reference data (e.g. titles, author names, venues, DOIs) against external publication records (libraries, metadata aggregators), an effort that scales poorly when combined with an already time-consuming review workflow.
As a result, fabricated or hallucinated citations can pass unnoticed, undermining the integrity of the evaluation process. 

This paper outlines an effort to make this emerging issue visible. By analyzing papers published in computer science conferences, we quantify the prevalence of "mysterious" citations, or those in which the cited title is not found at the cited location and also cannot be located through Google searches.  Rather than treating these incidents as isolated anecdotes, we provide empirical evidence of their scope and correlation to the advent of large language models, offering a foundation for informed discussion and policy in an era where generative models are a common part of academic authorship.

\section{USE AND MISUSE OF LLMS IN ACADEMIC WRITING}


Use of generative models in writing is not intrinsically problematic. There is a broad consensus that regular tasks such as grammar correction, formatting, and stylistic polishing do not, in themselves, threaten the foundations of academic communication. These uses can be considered similar to spellcheckers, reference managers, or automated typesetting tools, operating at the level of presentation rather than substance. 

The scientific method, by contrast, relies on transparent interaction with existing knowledge. Claims must be justified by evidence, and citations serve as verifiable links in a chain of progress. Generating invented sources breaks this chain. In other words, the unverified insertion of references that do not correspond to real works amounts to a form of citation fabrication, an operation indistinguishable from falsification, one of IEEE's and ACM's recognized categories of scientific misconduct~\cite{acm-content-falsification-policy-2025, ieee-ethics-fabrication-falsification-citation-2025}.

This distinction matters due to LLMs no longer being a fringe tool. Large-scale empirical work has shown that LLMs are already embedded in mainstream scientific writing practices. Liang et al.~\cite{liang2025quantifying} analyzed more than one million pre-prints and published articles and found a significant and rapidly increasing presence of LLM-modified text across disciplines with particular double digit adoption numbers in computer science. Their results provide clear evidence that generative models are not just experimental aids but are actively reshaping how scientific work is produced. 

Along with this broad adoption, however, a specific and consequential failure mode has begun to surface: the fabrication or hallucination of references. This pattern mirrors earlier high-profile incident outside of academia, where a legal brief drafted with LLM assistance was submitted containing judicial opinions and citations to cases that did not exist~\cite{weiser2023chatgpt}. The error was not stylistic, it produced plausible authority where none existed.

Comparable cases are now emerging in scientific publishing. Investigations of submissions to the 2026 International Conference on Learning Representations (ICLR) reported that 20\% (60 out of 300 papers sampled from 20,000 total submissions), contained at least one AI hallucination~\cite{esau2025gptzero}. GPTZero~\cite{tian2026gptzero} ``Hallucination Check'' scan found that 50 peer-reviewed submissions to ICLR contained at least one hallucinated citation that peer reviewers had not flagged. More recently, GPTZero found over 100 hallucinated citations in published NeurIPS papers~\cite{shmatko2026hallucinatedcitations}.  These fake references often include made-up authors, incorrect venues, or entirely invented citations~\cite{martin2025hallucinatedcitations}. At the time of writing, however, this citation-verification tool is not publicly available to authors or reviewers. 

While this paper was under review, several arXiv preprints identified hallucinated citations in published papers spanning a range of STEM disciplines~\cite{hallucitation, ghostcite}, highlighting the problem not only across computer science but also within other fields.

By situating LLM-assisted writing within both its legitimate uses and its emerging issues, we hope to underscore a central tension: generative models can responsibly assist with how academic writing is done, but if they are allowed to invent what they cite, they undermine the very mechanism by which scientific knowledge is validated and extended.

\section{METHODOLOGY}

Our experiments examine a single question: how has the prevalence of \textit{substantially erroneous} bibliographic entries changed since LLMs became publicly available.

This is distinct from the task of determining how prevalent LLM use is in the academic writing process \textit{more broadly}.
In general, deciding whether a given piece of text is generated by an LLM used by an adversarial author is an unsolved problem, and may even be impossible.
Even if it were possible, it would not be directly useful to determining the scientific validity of that text.
For example, a researcher acting in good faith could carry out a valid experiment, generate scientifically-grounded bullet points describing their work, use an LLM to produce a paragraph of prose, checked the resulting text for accuracy, and submit it for publication.
The mere fact that an LLM emitted the text is unrelated to the scientific validity.

Unlike other prose, scholarly writing feature structured references to preexisting related publications.
To a much greater extent than other parts of prose, the references themselves can be checked for correctness.
This is not a judgment of the merit of the referenced work or it's relationship to the text, but rather a straightforward check of whether the reference points to an extant, preexisting work.

These structured references provide an opportunity to examine the prevalence of a specific pattern of LLM misuse: generating related bibliography entries without actually consulting the related work in question.
It is difficult to defend this use of LLMs as being part of a valid scholarly process, separate from other potential uses of LLMs.

\begin{figure*}
\centering
\includegraphics[width=\textwidth]{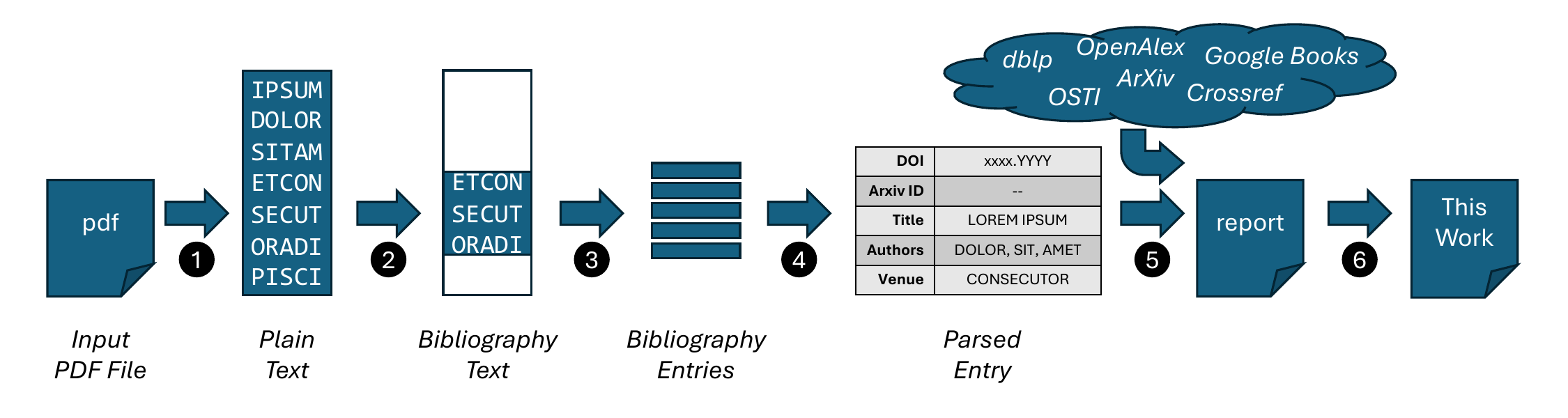}
\caption{Summary of the analysis methodology.
\ding{202} extract text via PyPDF's \texttt{ParsePDF}.
\ding{203} isolate bibliography.
\ding{204} split bibliography into entries.
\ding{205} parse bibliography entry.
\ding{206} search external aggregators for matching publications.
\ding{207} manually validate machine results.
}
\label{fig:pipeline}
\vspace*{-5pt}
\end{figure*}

Published papers often have 30 to over 100 citations, creating large barriers to manually verifying their correctness.  To validate references within a published paper, the PDF file can be passed through a software tool~\cite{bibcheck}, which produces a summary report (\ding{202}-\ding{206} in Figure).  The report flags all citations for which a perfect match is not found in online metadata.  All flagged citations must be manually checked. 

On average, the software tool reduces the number of citations that must be manually checked to around 5 to 10.  There are many reasons non-mysterious citations may need manually checked, including citations to prior versions of ArXiv pre-prints as well as USENIX publications, which are not consistently stored within the searched metadata.  Note, this pipeline was created to ensure the correctness of citations within a single publication, such as by an author, reviewer, or reader.  No AI is used within the pipeline to avoid false negatives while also providing a tool for reviewers, who are often forbidden from using generative AI.  There is a significant amount of manual overhead required to validate entire proceedings with this pipeline. 
Figure~\ref{fig:pipeline} summarizes this process.

\ding{202} \textbf{Extract PDF text}: PyPDF's~\cite{pypdf} \texttt{ParsePDF} method is used to convert the input PDF into plain text.

\ding{203} \textbf{Isolate bibliography}: String search heuristics are used to isolate the paper's bibliography; specifically, the text between the last occurrence of ``bibliography'' or ``references'' and the following occurrence of ``appendix'' (or the end of the document).

\ding{204} \textbf{Split bibliography into entries}: String search heuristics are used split the bibliography into separate entries. For the conferences analyzed in this work, all bibliography entries started with a number in square brackets (e.g. ``[17]'').

\ding{205} \textbf{Parse entry}: String search is used to find DOIs and ArXiv identifiers. Regular expressions are used to identify title, authors, and venue from 12 empirically-identified reference formats.

\ding{206} \textbf{Search external aggregators for matching publications}: Depending on the extracted metadata, a variety of external sources are searched for matching publications. Those sources are ArXiv, Crossref, dblp, Google Books, OpenAlex, and the U.S. Department of Energy Office of Science and Technical Information (OSTI).  Matches are determined with Levenshtein ratio.  To prevent false negatives, we only consider metadata a match for a ratio of 1.

\ding{207} \textbf{Manually validate machine results}: If a citation is not found automatically, the pipeline prints the closest match while highlighting differences.  For each flagged citation, a comparison must done with the closest match to check for typos.  Assuming no similar title or authors were found, a manual search for the article must be carried out. If a web search does not validate the existence of the cited publication, the cited pages within the journal volume or proceedings are checked for the cited content. For ArXiv pre-prints, previous versions of the pre-print are checked for title and author discrepancies.

\section{MYSTERIOUS CITATIONS}
We have chosen to anonymize all presented results to avoid implicating any specific individuals, institutions, or venues.  Rather, than focusing on specific conferences, we hope to raise awareness of this systemic challenge to the scholarly writing process.

We analyzed the proceedings of four high-performance computing conferences for mysterious citations and incorrect author lists.  All examined conferences are well-known and respected within the community and were selected due to our familiarity as previous authors or program committee members.  These conferences, labeled as Conferences 0 through 3 throughout the remainder of this section, have ERA conference rankings of A, C, A, and non-indexed, respectively~\cite{conf-ranks}. The 2021 proceedings were analyzed as a baseline before generative AI while the 2025 proceedings of the same conferences were examined for current trends.  We analyzed all papers within each proceeding for correctness of both titles and authors. Note that the proceedings vary in size, from less than 50 to over 100 publications.

Citation errors naturally can occur for a number of reasons, including typos and discrepancies among subtitle inclusion.  Therefore, we have ignored all minor errors, such as misspelled words or a small number of missing words.  We categorize the remaining mysterious citations into two groups:
\begin{enumerate}
    \item \textbf{Rephrased Title:} The title has been rephrased but retains the meaning of the true title.  A similar publication is found at the listed location with the listed authors.\\
    \item \textbf{Fully Mysterious Citation:} The cited location either does not exist or holds an unrelated paper with different authors.  No similar publication is found through Google searches.
\end{enumerate}
\begin{figure}[ht!]
    \centering
    \includegraphics[width=0.9\linewidth]{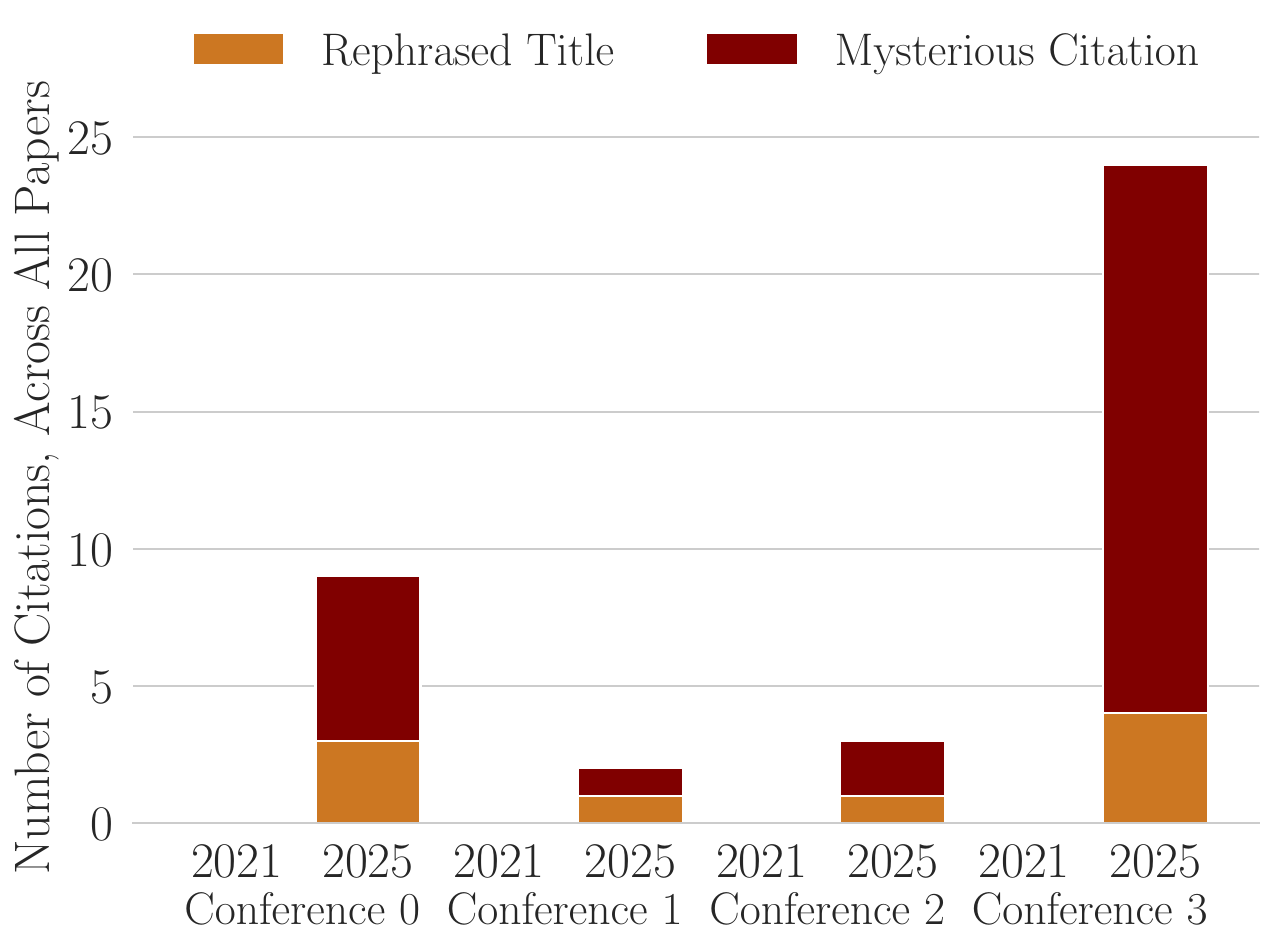}
    \caption{Baseline comparison: number of rephrased titles and fully mysterious citations, per conference, per year. No occurrences were observed in 2021.}
    \label{fig:title_baseline}
\end{figure}
Figure~\ref{fig:title_baseline} compares citation errors within the 2021 and 2025 proceedings of all four conferences.  While no mysterious citations were found in any of the 2021 proceedings, both rephrased titles and fully mysterious citations were found in every one of the 2025 proceedings, correlating with the advent of generative AI.

\begin{figure}[ht!]
    \centering
    \includegraphics[width=0.49\textwidth]{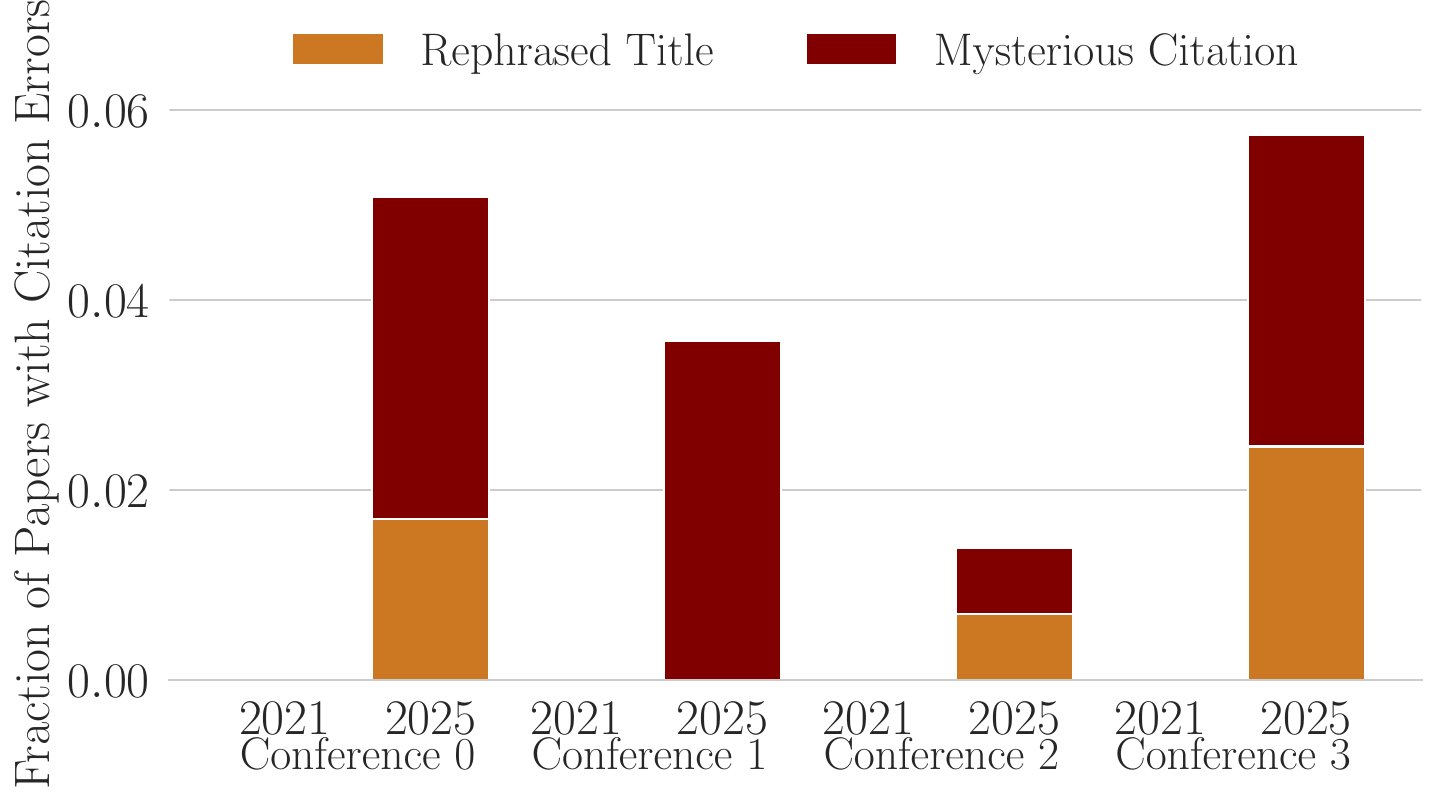}
    \caption{The fraction of papers with citation errors.  If any citation within a paper is fully mysterious, the paper is colored maroon, while orange papers represent those that contain only rephrased titles.  No occurrences were observed in 2021.}
    \label{fig:errors2025}
\end{figure}
Figure~\ref{fig:errors2025} shows the percentage of papers with mysterious citations across the proceedings.  Both rephrased titles and fully mysterious citations were found in all four conference proceedings, with the fraction of papers in which they were found ranging from under 2\% of papers to nearly 6\%.  \textbf{In every proceedings that we examined, there are \emph{published} papers that include citations to papers for which we were able to find no evidence of existence.}  In most of the papers in which a mysterious citation was found, one or a small number of citations were found to be mysterious.  However, in one paper, over half of the citations contained either rephrased titles or were fully mysterious.  \textbf{In total, 13 of the 353 published papers were found to contain a total of 38 mysterious citations.}  Five of these papers were published in the two analyzed A-ranked conferences.

\subsection{Author and DOI Errors}
In addition to the mysterious citations, we also analyzed errors within author lists.  Author name misspellings, reversals of first and last names, missing authors, and extra authors were frequently found across all proceedings.  Therefore, we examined recent proceedings only for author lists in which the majority of listed authors were incorrect.  Citations were only included in this analysis if more than 50\% of the cited authors were not listed on the cited publication.  In the case of fully mysterious citations, all authors were considered incorrect. 
\begin{figure}[ht!]
    \centering    \includegraphics[width=1.0\linewidth]{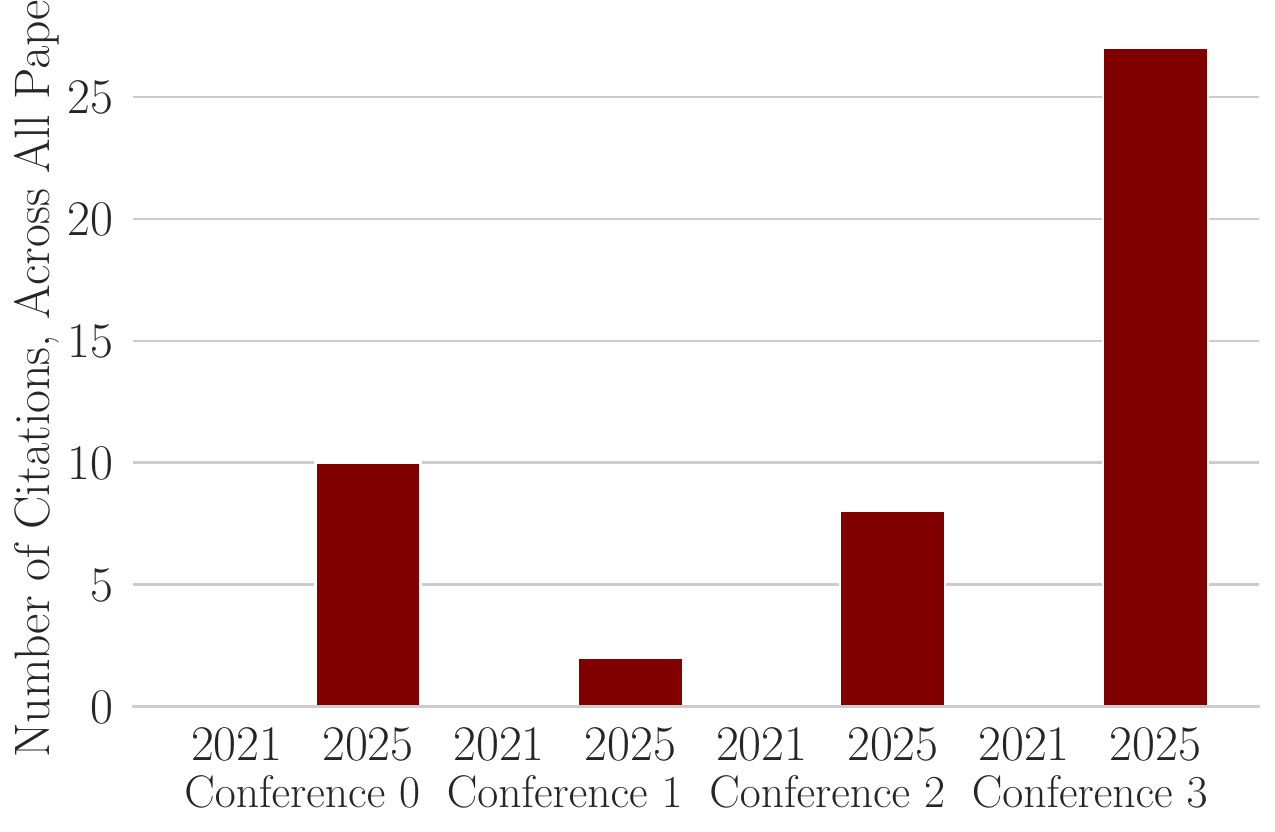}
    \caption{Baseline comparison: number of citations, per conference, in which over half of the cited authors were not listed on the cited paper. No occurrences were observed in 2021.}
    \label{fig:authors}
\end{figure}
Figure~\ref{fig:authors} shows the number of citations, per conference, in which the majority of cited authors were incorrect. While there were no occurrences in any analyzed 2021 proceeding, majority incorrect author lists were found in multiple citations in every 2025 proceedings.  \textbf{In total, 15 of the 353 published papers included citations in which over half of the listed authors were incorrect, for a total of $46$ occurrences.}  Eight of these papers were published in the two analyzed A-ranked conferences.

Finally, we analyzed the correctness of provided DOIs and Arxiv IDs.  The vast majority of citations did not include a DOI or Arxiv ID.  While there were incorrect DOIs and ArXiv links in multiple of the examined proceedings, both in 2021 and 2025, trends are difficult to analyze due to the small sample size.

\subsection{AI-Generated Citations}
There is a correlation between the introduction of generative AI and the occurrence of mysterious citations within publications.  However, we are unable to verify that AI was used to generate the mysterious citations found in this work.  \textbf{All analyzed conferences required authors to acknowledge any generative AI usage, yet none of the papers acknowledged using AI to generate citations.}  During our data collection, we did find one non-mysterious citation that included a URL ending in \texttt{utm\_source=chatgpt.com}, strongly indicating the citation was generated by ChatGPT.  However, the authors only acknowledged using generative AI for grammatical edits.  

\begin{figure}[ht!]
    \centering
    \includegraphics[width=0.8\linewidth]{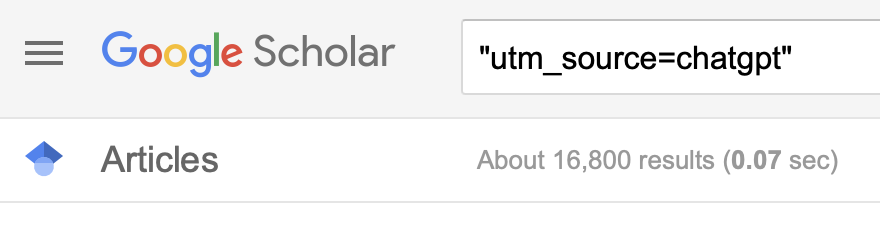}
    \caption{A Google Scholar search for articles containing URLs generated by ChatGPT, showing 17,000 results.}
    \label{fig:utm_source}
\end{figure}

While we limited our dataset to proceedings of HPC conferences with which we are familiar, AI-generated citations are appearing across research domains. For instance, the Google Scholar search displayed in Figure~\ref{fig:utm_source} shows that \textbf{at the time of writing this article, nearly 17,000 articles contain a link generated by ChatGPT}.

\subsection{AI-Generated Papers}
Without author acknowledgment, there is no way to verify which portions of a paper have been generated by AI.  However, in recent years, non-peer-reviewed papers have been published online by unverified researchers at a rapid rate.  This is particularly problematic on ResearchGate~\cite{researchgate}, where no verification process is required to create a profile and upload a paper, enabling fake online personas such as Larry, the world's most cited cat~\cite{larry}.  

One of the citations from Conference 3 was verified as fully mysterious, with no related article existing in the cited location. However, while the exact title was not found in any peer-reviewed publication, it did appear on ResearchGate.  
\begin{figure}[ht!]
    \centering
    \includegraphics[width=1.0\linewidth]{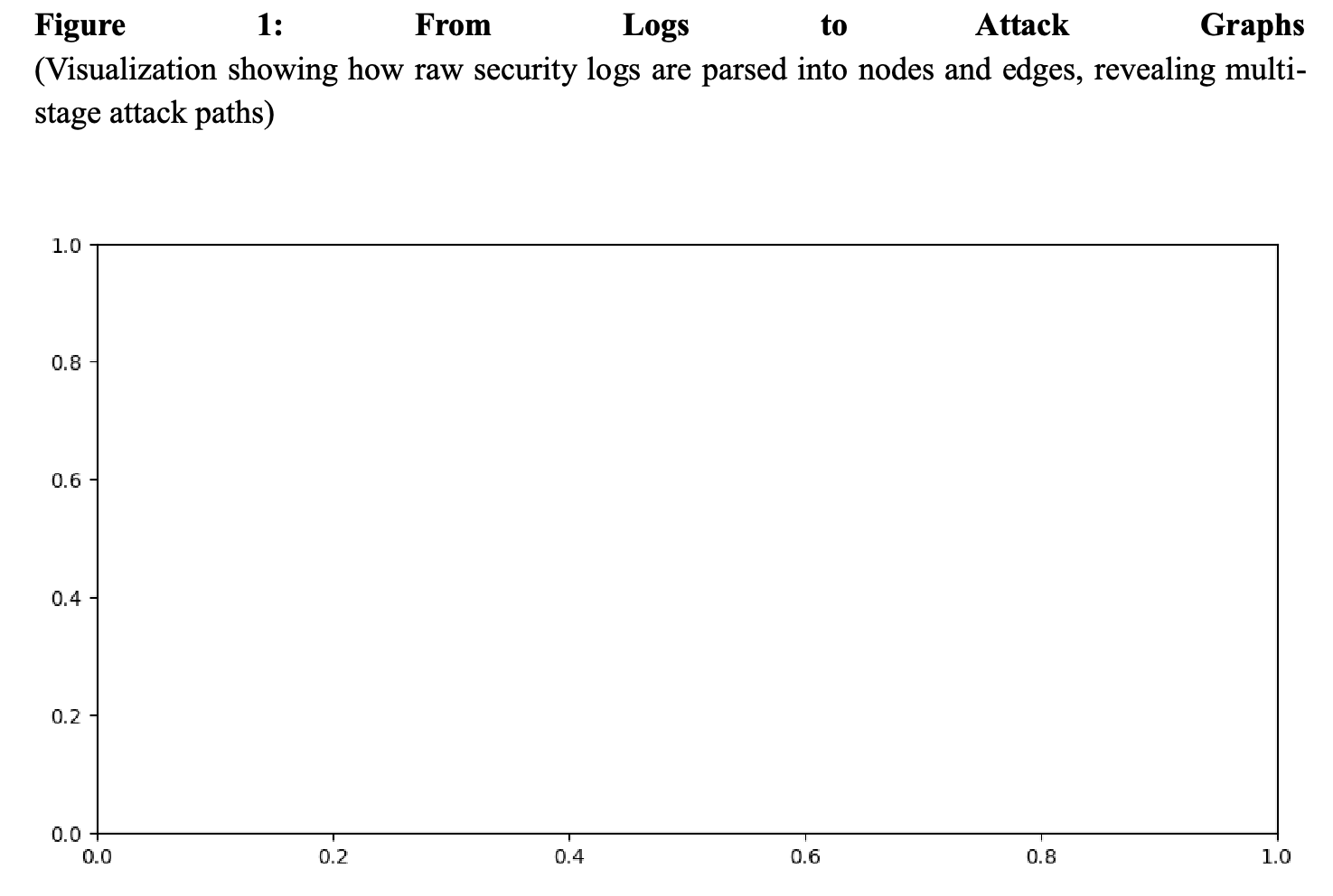}
    \caption{A screenshot of the data presented in the ResearchGate upload cited by a paper in Conference 3.}
    \label{fig:fake_paper}
\end{figure}
The full text was uploaded, but it was not formatted for formal publication, contained bulleted lists instead of paragraphs, and had empty figures as shown in Figure~\ref{fig:fake_paper}.  \textbf{This incomplete paper was uploaded to ResearchGate on the same date as the citing paper's submission deadline, as listed in Conference 3's Call for Papers.}  The ResearchGate profile to which this paper was uploaded has since added 40 similar papers.

\section{DISCUSSION}
Inappropriate and unacknowledged use of generative AI poses a growing risk to the integrity of the scientific community.  When a paper contains unacknowledged generated content, such as embedded \texttt{utm\_source=chatgpt}, the community is left to infer that generative AI may have been used in other unacknowledged ways.  More concerning, when a paper includes mysterious citations, readers are left to assume the authors either intentionally fabricated references or included unverified generated content throughout the publication.  Such issues can undermine trust in the work as a whole and can negatively impact the reputation of all co-authors, even those who were unaware generated content was included.

Crucially, every mysterious citation identified in this study appears in published proceedings. These papers successfully passed peer review and editorial checks, yet still contain fabricated or corrupted references. This demonstrates that the current review process which is already under severe time and workload constraints, is not sufficient to reliably detect these classes of errors.

The scientific community could take steps to better prevent these mysterious citations from occurring within our peer-reviewed publications.  If an author uses generative AI during a literature review, hallucinations can be avoided by manually finding and reading the publication, using only official resources to cite it.  Co-authors can prevent mysterious citations by checking the full bibliography for correctness before submission.  

During the review process, all citations for each article can be verified prior to acceptance. Conference organizing committees can further develop policies, such as desk rejecting papers with mysterious citations or/and treat fake citations as a form of scientific misconduct on par with fabrication. Major publication venues could adopt an explicit policy that defines unacceptable citation practices and associated penalties. For difficult to find references, authors could be required to supply verifiable evidence such as DOI or PDF at submission time. 

Fully hallucinated citations are likely to disappear as generative AI continues to improve.  To date, while citations generated by ChatGPT 5 consistently include extra words, incorrect authors, and invalid DOIs, all of our attempts to prompt the new model into generating a fully fabricated citation have been unsuccessful.  However, hallucinated citations are only a tangible symptom of a larger underlying problem.  Researchers are increasingly using AI to generate content and, in some cases, publishing the generated text without verifying it for correctness.  This trend raises questions about whether research papers can remain a reliable method for sharing scientific progress.

\section{CONCLUSION}
Our study provides the first systemic evidence that mysterious citations have begun to regularly appear in peer-reviewed conferences proceedings. By comparing four major venues from 2021 and 2025, we show that every 2025 proceeding we examined contains at least one mysterious citation. These errors, which affect up to 6\% of \textit{published papers} in some venues, pose a serious issue for the integrity of academic communication and are unlikely to be caught by today's review processes. In order to initiate community discussion on this topic, we briefly mention some actions that could be taken to mitigate mysterious citation errors. Combining individual diligence with explicit community-wide policy on this subject and developing automated tools could help improve traceability and maintain trust in the scientific process.



\section{LIMITATIONS}
It is not possible to prove that a mysterious citation was generated through an LLM hallucination.  After hallucinated citations are published, they may be added into metadata, leading to authors acquiring mysterious citations from trusted sources such as Google Scholar~\cite{hallucitation}.  However, the problem remains that authors are adding unverified data into their papers.  

It is possible that additional significant citation errors exist in the analyzed proceedings and were not caught as a part of this analysis.  For example, if the error originated in online metadata, a match for the erroneous title would be found by the tool.  Further, this paper only reports mysterious citations that could be verified as incorrect, in which the cited journal or proceeding could be searched to verify that the citation does not exist.  The pipeline has no way to catch mysterious citations to websites or blogs as there is no way to prove the cited location did not exist at one time.

\section{ACKNOWLEDGMENTS}
This work has been submitted to the IEEE for possible publication. Copyright may be transferred without notice, after which this version may no longer be accessible.

This work made use of free, public API access to ArXiv, Crossref, dblp, OpenAlex, and the U.S. Department of Energy Office of Science and Technical Information (OSTI). The authors would like to thank those organizations for providing those resources.  Generative AI was used to generate string parsing as API search keys when developing the citation verification tool.

This work was performed with partial support from the National Science Foundation under Grant Nos. CCF-2338077, CNS-2450092, and OIA-2521103 and the U.S. Department of Energy's National Nuclear Security Administration (NNSA) under the Predictive Science Academic Alliance Program Awards DE-NA0003966, DE-NA0004267.

Any opinions, findings, and conclusions or recommendations expressed in this material are those of the authors and do not necessarily reflect the views of the National Science Foundation and the U.S. Department of Energy's National Nuclear Security Administration.

Sandia National Laboratories is a multi-mission laboratory managed and operated by National Technology \&  Engineering Solutions of Sandia, LLC (NTESS), a wholly owned subsidiary of Honeywell International Inc., for the U.S. Department of Energy’s National Nuclear Security Administration (DOE/NNSA) under contract DE-NA0003525. This written work is authored by an employee of NTESS. The employee, not NTESS, owns the right, title and interest in and to the written work and is responsible for its contents. Any subjective views or opinions that might be expressed in the written work do not necessarily represent the views of the U.S. Government. The publisher acknowledges that the U.S. Government retains a non-exclusive, paid-up, irrevocable, world-wide license to publish or reproduce the published form of this written work or allow others to do so, for U.S. Government purposes. The DOE will provide public access to results of federally sponsored research in accordance with the DOE Public Access Plan.

\def\refname{REFERENCES}
\bibliographystyle{ieeetr}  
\bibliography{refs.bib}


\begin{IEEEbiography}{Amanda Bienz}{\,} is an Assistant Professor in the Computer Science Department at the University of New Mexico. Her research interests include improving the performance and scalability of parallel solvers, applications, and MPI collective operations for emerging architectures.  Bienz received her Ph.D in Computer Science from the University of Illinois Urbana–Champaign.  Contact her at bienz@unm.edu.\vspace*{8pt}
\end{IEEEbiography}

\begin{IEEEbiography}{Carl Pearson}{\,} is a Senior Member of Technical Staff in the Scalable Algorithms group at Sandia National Laboratories. His research interests include future architectures for scientific computing, GPU communication for distributed linear algebra, and GPU acceleration of irregular operations. He received his Ph.D in Electrical and Computer Engineering from the University of Illinois Urbana–Champaign. Contact him at cwpears@sandia.gov.\vspace*{8pt}
\end{IEEEbiography}

\begin{IEEEbiography}{Simon Garcia de Gonzalo} {\,} is a Senior Member of Technical Staff in the Scalable Computer Architecture group at Sandia National Laboratories. His research interests include high-performance computing architectures, GPU communication and benchmarking, and performance analysis for emerging heterogeneous systems. He received his Ph.D. in Computer Science from the University of Illinois Urbana–Champaign. Contact him at simgarc@sandia.gov.
\end{IEEEbiography}

\end{document}